\documentstyle[preprint,axodraw,aps]{revtex}

\begin{document}

\title{Relativistic dynamics of Qqq systems 
\footnote{To Appear Phy. Rev. D (2002).}}
\author{ E.F. Suisso$^a$, J. P. B. C. de Melo$^b$ and 
T. Frederico$^a$}
\address{\
$^a$ Dep. de F\'\i sica, Instituto Tecnol\'ogico da Aeron\'autica, \\
Centro T\'ecnico Aeroespacial, 
\\ 12.228-900, S\~ao Jos\'e dos Campos, SP, Brazil\\
$^b$Instituto de F\'\i sica Te\'orica, Universidade Estadual Paulista \\
0145-900, S\~ao Paulo, SP, Brazil
}
\date{\today}
\maketitle
\begin{abstract}
The bound state of constituent quarks
forming a $Qqq$  composite baryon is investigated in a
QCD-inspired effective light-front model.
The light-front Faddeev equations 
 are derived and solved numerically. The masses of the
spin 1/2 low-lying states of the nucleon, $\Lambda^0$, 
$\Lambda^+_c$ and $\Lambda^0_b$ are found and compared 
to the  experimental data. The data  is
qualitatively described with a flavor independent effective
interaction.
\end{abstract}
\section{Introduction}
One major  task in strong interaction physics is the calculation 
of the wave function and the spectrum of the hadrons
from Quantum Chromodynamics (QCD)\cite{iz}. Phenomenological
dynamical models that retain the low-energy physics of QCD, and
relate different observables are still of interest as long as they
allow to expose the nonperturbative  properties of QCD.
 One possibility to model 
the quark dynamics within a 
relativistic framework  is to use the light-front dynamics 
in a truncated Fock-space, which yields a wave-function 
covariant under kinematical boosts \cite{brodsky,karmanov}.
In general, the light-front Fock-state truncation is stable
under kinematical boost transformations\cite{perry}. 

A light-front QCD-inspired model was recently applied to
the pion and other mesons\cite{pauli,tobpauli}. It was able
to describe the pion structure as well as the masses of the vector
and pseudo-scalar mesons. This effective model, 
the $\uparrow\downarrow$-model\cite{pauli}, has two components
in the interaction, a contact term and a Coulomb-type potential. The
contact term is essential to collapse the constituent quark-antiquark system
to form the pion, while the vector meson is dominated
by the Coulomb-type potential. In this model the vector meson corresponds 
to a weakly bound system of constituent quarks. The model
has no confinement, and the spin does not play a dynamical role
besides justifying the contact term from the hyperfine interaction.
However, the contact term was able to embrace the
physical scale brought by the pion mass and from that the masses of the
other pseudo-scalar and vector mesons were calculated\cite{tobpauli}.
The flavor independence of the interaction was assumed, 
and the model reproduced fairly well the data, despite its simplicity.

Here we report a nonperturbative calculation of the flavor dependence 
of the masses of some baryons, extending the concepts coming from the 
$\uparrow\downarrow$-model applied previously only to
mesons\cite{pauli,tobpauli}.
In this work, we show that the flavor independence of the effective
interaction still holds for baryons, by studying the
spin 1/2 low-lying states of the nucleon, $\Lambda^0$, 
$\Lambda^+_c$ and $\Lambda^0_b$. To get insight on the complex
three-quark relativistic dynamics in this 
first study of the $Qqq$ system within the light-front framework, we use
only the contact interaction, which brings the physical scale of the 
ground state of the nucleon, while the spin is averaged out. 
The effect of the long-range Coulomb-type interaction is effectively
carried out by tunning the contact interaction to the nucleon mass.
The mass of one of the constituent quarks $(Q)$ will be varied 
while the bare strength of the effective contact interaction is kept 
constant. For each constituent mass  the binding energy
of the three-quark system is evaluated. Naturally, this 
calculation yields the binding energy of the constituent 
quarks as a function of the baryon ground state mass, 
because the binding energy and the ground state mass,
depend only on the mass of the quark $Q$, with the other
inputs kept unchanged.

The binding energy for constituent quarks is a difficult concept to  use
together with quark confinement which is believed to  exist in nature. 
It is fair to ask, if one has a meaningful model without confinement, 
how one could extract from the experimental data a quantity to be 
compared with the model binding energy. This key point will be addressed
in detail in the text. 

Let us mention that, the three-body model with constituent quarks interacting 
in the light-front with a contact force\cite{tob92}  has been applied 
to the proton and  described its mass, charge radius and electric 
form factor up to 2(GeV/c)$^2$\cite{Pacheco95},
although the spin was averaged out and before the advent of the 
$\uparrow\downarrow$-model\cite{pauli,pauli98}. Recently the same
three-quark model
was applied to study the dissolution of the nucleon at 
finite temperature and baryonic density\cite{beyer01}.

The paper is organized as follows. In Sec.II, we derive the coupled 
integral equations for the Faddeev components of the vertex of the
 three-body light-front bound-state wave function\cite{tob92},  
 which generalizes the Weinberg-type integral equation found
 for a two-body bound state\cite{Weinberg66}. In Sec. III,
 we present the numerical results for the masses of the nucleon,
$\Lambda^0$, $\Lambda^+_c$ and $\Lambda^0_b$, using the flavor
independent contact interaction. We show as well how to relate
to experimental data the results for the binding energies. 
Finally, in Sec.IV, we give our conclusions.

\section{\bf THREE-QUARK RELATIVISTIC MODEL}

The light-front  is  defined by $x^{+}=x^{0}+x^{3}=0$ and the 
coordinates in this space-time hypersurface are given by 
$x^{-} = x^{0}+x^{3}$ and $\vec{x}_{\perp}=(x^1,x^2)$\cite{brodsky,karmanov}. 
The coordinate  $x^{+}$ is recognized as the time and $k^{-}= k^{0}-k^{3}$, 
the momentum canonically conjugated, corresponds to the light-front energy. 
The momentum coordinates $k^{+}$ and 
$\vec{k}_{\perp}$, are the kinematical momenta canonically
conjugated to $x^{-}$ and $\vec{x}_{\perp}$, respectively.

A relativistic model for three-particles on the light-front
for a pairwise contact interaction, was derived from the
three-body ladder Bethe-Salpeter equation for the Faddeev component of the
vertex function, by eliminating the
relative $x^+$-time between the particles\cite{tob92}. 
The projection of the covariant dynamics to the light-front hypersurface,
is performed through the integration over the $k^-$ momentum of the
individual particles in the Bethe-Salpeter equation
leading to a Weinberg-type equation\cite{Weinberg66} for three-particles. 
In the work of Ref. \cite{tob92}, in fact the kernel of the
Faddeev equation in the light-front was derived in lowest order, and 
in principle corrections of higher order can be systematically constructed
following  recent discussions\cite{sales00,miller,brinet}.                  

Let us sketch the derivation of the light-front Faddeev equations
for a heavy-light-light three quark system ($Qqq$) from the four-dimensional
Bethe-Salpeter equations. Two different
spectator functions, which corresponds to the Faddeev components of the
vertex, are possible. For the interacting pair being $qq$ the
spectator function is $v_{Q} (q^\mu)$, function of
the four-vector momentum of the quark $Q$.
For the interacting pair $Qq$ the spectator function is
$v_{q} (q^\mu)$. The coupled Faddeev-Bethe-Salpeter equations in the
ladder approximation are given by: 
\begin{eqnarray}
v_Q\left( q^\mu \right) &=&-2i\tau _{qq }\left( M_{qq}^2\right)
\int \frac{d^4k}{(2\pi)^4}\frac{v_q\left( k^{\mu }\right) }{(k^2-
m_q^2+\imath \epsilon)((P_B-q-k)^2-m_q^2+\imath \epsilon)}\ ,
\label{bs2} 
\end{eqnarray}
which is represented diagrammatically in figure 1, and
\begin{eqnarray}
v_q\left( q^\mu \right) =-i\tau _{Qq}\left( M_{Qq}^2\right)
\int \frac{d^4k}{(2\pi)^4}&&\left[\frac{v_q\left( k^\mu \right)}{(k^2-
m_q^2+\imath \epsilon)((P_B-q-k)^2-m_Q^2+\imath \epsilon)} \right.
\nonumber \\ && 
\left. + \frac{v_Q\left( k^\mu \right) }{(k^2-
m_Q^2+\imath \epsilon)((P_B-q-k)^2-m_q^2+\imath \epsilon)} \right],
\label{bs3} 
\end{eqnarray}
which is represented in figure 2. The baryon four-momentum is
given by $P_B$, the light and heavy quark masses are
$m_q$ and $m_Q$, respectively. The masses of the
virtual two-quark subsystems are $ M_{qq}^2=(P_B-q)^2$ and
$ M_{Qq}^2=(P_B-q)^2$ due to the conservation of the total four-momentum.
The two-quark scattering amplitudes 
$\tau _{qq}\left(M_{qq}^{2}\right)$ and
$\tau _{Qq }\left( M_{Qq}^{2}\right)$
are the solutions of the Bethe-Salpeter equation in the 
ladder approximation  for a contact interaction between 
the quark pairs, which are derived in detail in the appendix.

The analytical integration 
over $k^-$ is performed in Eqs. (\ref{bs2}) and (\ref{bs3}), using only
the pole of the single quark propagator
in the lowest half of the complex $k^-$ plane.
The pole is given by the on-energy-shell
condition $k^-_{on}=(k^2_\perp+m^2_{\alpha})/k^+$, with $\alpha= \ q$ or $Q$
when $v_q\left( q^\mu \right)$ or $v_Q\left( q^\mu \right)$
is integrated, respectively. The condition for a nonvanishing result of the
integration in $k^-$ is $0 \ <\  k^+\ <  \ P^+_B -q^+$. The spectator
functions appearing inside the integrations in Eqs. (\ref{bs2})
and (\ref{bs3}) depend only on the kinematical momentum, 
($k^+,\vec k_\perp$), as long as $k^-_{on}$ is a function of the kinematical
momentum. Thus, to close the light-front equations the external
momentum is chosen on the $k^-$-shell.

In the rest-frame of the baryon of mass $M_B$, we write that 
$v_{\alpha} ( q^+,\vec{q}_{\perp},q^-_{on})\equiv 
v_{\alpha} ( \vec{q}_{\perp},y )$, where for
convenience the Bjorken momentum fraction $y=q^+/M_B$ is used. 
The Faddeev equations in the light-front  written in terms of the kinematical 
momenta are given by:
\begin{eqnarray}
v_{Q} ( \vec{q}_\perp,y ) =&&-\frac{2i}{\left( 2\pi
\right) ^3}\tau _{qq }\left( M_{qq}^{2}\right)
\int\limits_{0}^{1-y}\frac{dx}{x\left( 1-x-y\right) }
\int d^{2}k_{\perp }
\nonumber \\
&&\times
\frac{\theta \left( x-\frac{m_q^2}{M^2_B}\right) 
\theta\left( k_\perp^{\max}(m_q)-k_\perp \right)}{M_B^2-
\frac{q_\perp^2+m_Q^2}{y}-\frac{k_\perp^2+m_q^2}{x}-
\frac{\left(P_B-q-k\right)_\perp^2+m_q^2}{1-x-y}}
v_{q}(\vec{k}_\perp,x) ,
\label{Mass3}
\end{eqnarray}
\begin{eqnarray}
v_{q} ( \vec{q}_{\perp},y ) &=&-\frac{i}{\left( 2\pi
\right) ^3}\tau _{Qq }\left( M_{Qq}^{2}\right)
\int\limits_{0}^{1-y}\frac{dx}{x\left(1-x-y\right)}
\int d^{2}k_{\perp }%
 \nonumber \\
&&\times \left[
\frac{
\theta \left( x-\frac{m_q^2}{M^2_B}\right) 
\theta\left( k_\perp^{\max}(m_q)-k_\perp\right)}{M_B^2
-\frac{q_\perp^2+m_q^2}{y}-\frac{k_\perp^2+m_q^2}{x}
-\frac{\left(P_{B}-q-k\right) _{\perp }^2+m_Q^2}{ 1-x-y }}
v_{q}(\vec{k}_\perp,x)
\right.
\nonumber \\
&&\left. +
\frac{
\theta \left( x-\frac{m_Q^2}{M^2_B}\right) 
\theta\left( k_\perp^{\max} (m_Q)-k_\perp\right)}
{M_B^2-\frac{q_\perp^2+m_q^2}{y}
-\frac{k_\perp^2+m_Q^2}{x}-
\frac{\left(P_{B}-q-k\right)_\perp^2+m_q^2}{1-x-y}} 
v_{Q}(\vec k_\perp,x)
\right] \ ,
\label{Mass32}
\end{eqnarray}
where $x=k^+/M_B$.
In the first equation of the coupled set, Eq. (\ref{Mass3}), $Q$ 
is the spectator quark and $qq$ is the interacting pair.
The maximum value for $k_{\perp}$ is chosen to keep the mass squared
of the $qq$ or $Qq$ subsystem real, i.e., $M^2_{qq} \ge 0$
 and $M^2_{Qq} \ge 0$, respectively. These constraints in the
spectator quark phase-space  come through the theta functions
in the integrations of Eqs. (\ref{Mass3}) and 
(\ref{Mass32}). For $M^2_{Qq}\ge 0$ one has
$k_\perp<k_{\perp}^{\max }( m_{q}) =\sqrt{( 1-x) (
M_{B}^{2}x-m_{q}^{2})}$, and $x\ge (m_q/M_B)^2$.
For $M^2_{qq}\ge 0$ one has
$k_\perp<k_{\perp}^{\max }( m_{Q}) =\sqrt{( 1-x) (
M_{B}^{2}x-m_{Q}^{2})}$, and $x\ge (m_Q/M_B)^2$.
For equal particles, Eq.(\ref{Mass3}), reduces to the one
derived in Ref. \cite{tob92}. 

Finally, the light-front baryon bound state wave-function of
the $Qqq$ system in the rest-frame is constructed from the
Faddeev components of the vertex as in Ref.\cite{Pacheco95}:
\begin{equation}
\Psi ( x_{1},\vec{k}_{1\perp };x_{2},
\vec{k}_{2\perp}) =
\frac{v_{q}\left( x_{1},\vec{k}_{1\perp }\right)
+v_{q}\left( x_{2},\vec{k}_{2\perp}\right) + 
v_{Q}\left( x_{3},
\vec{k}_{3\perp}\right) }{\sqrt{x_{1}x_{2}x_{3}}\left(
M_{B}^{2}-M_{0}^{2}\right)}  \ ,
\label{Mass31}
\end{equation}
where the free three-quark mass $M_0^2$ is given by:
\begin{equation}
M_0^2=\frac{k^2_{1\perp }+m_q^2}{x_{1}}+
\frac{k^2_{2\perp }+m_q^2}{x_{2}}+
\frac{k^2_{3\perp }+m_Q^2}{x_{3}} \ .
\label{Mass19}
\end{equation}
Each constituent  quark has momentum fraction $x_j$ and transverse momentum 
$\vec{k}_{j \perp}$$(j=1,3)$, satisfying $x_1+x_2+x_3=1$ and 
$\vec{k}_{1 \perp} + \vec{k}_{2 \perp} + \vec{k}_{3 \perp}=0$.

\section{\bf RESULTS} 

The coupled integral equations  (\ref{Mass3}) and (\ref{Mass32})
for a relativistic system of three constituent quarks with  
a pairwise zero range interaction, are solved numerically.
The physical inputs of the model are the constituent quark
masses and the diquark bound state mass.
 The mass of the ground state baryon $(M_B)$ and the  binding energy,
$ B_B=2m_q+m_Q-M_B$, are calculated. 
First in this section, before presenting the
model results, we provide a qualitative discussion in order to attribute
to the low-lying spin 1/2 baryons a binding energy from the 
experimental data. Then, we compare these data with the model
calculations.

\subsection{Qualitative analysis}

According to the effective QCD-inspired model calculations 
of Ref. \cite{tobpauli}, the low-lying vector mesons
are weakly bound systems of constituent quarks while the
pseudo-scalars are more strongly bound. This justifies our supposition 
that the masses  of the constituent quarks  can be derived directly from the 
vector meson masses:
\begin{eqnarray}
 m_u &=&\frac{1}{2}M_{\rho}=0.384{\text GeV}  \ ,\nonumber \\ 
 m_s&=&M_{K^{*}} - \frac{1}{2}M_{\rho}=0.508{\text GeV} \ , \nonumber \\
 m_c&=&M_{D^{*}} - \frac{1}{2}M_{\rho}= 1.623{\text GeV} \ , \nonumber \\
 m_b &=&M_{B^{*}} - \frac{1}{2}M_{\rho}=4.941{\text GeV}\ ,
\label{mconst}
\end{eqnarray}
where the values from Table I are used. Also one can check
whether the values of the current quark masses obtained from 
the constituent ones, attain values compatible with the
actual knowledge\cite{pdg}.
It is reasonable to think that the constituent quark mass formation 
mechanism does not distinguish in detail the quark flavor and thus,
the current quark mass of $s$, $c$ and $b$ are just given by the 
difference $m^{curr}_Q=m_Q-m_u$ ($Q=s,c,b$).  
The extracted values of the current masses  are quite consistent with the 
experimental ones from Ref.\cite{pdg}, as one can verify in Table I.
At least from this point of view is not unaceptable to define the
constituent quark masses from the low-lying vector meson states.

A remark should be added on how ambiguous the qualitative estimates are for the
constituent and current quark masses. Errors arise when
different meson masses are used as input, and from the
current quark masses of the up and down quarks, 
which we have disregarded and leads to an error of about 10 MeV. 
Moreover, another 10 MeV can be attributed to the degeneracy 
of the $\rho$ and $\omega$ mesons in our model. 
Therefore, we roughly estimate an error of 20 MeV in Table I for
our values of current quark masses, and for the constituent quark masses from 
Eq.(\ref{mconst}), as well.

Below, we attribute values to the baryon binding energies, using 
the constituent quark masses from Eqs. (\ref{mconst}) and the
experimental values of the baryon masses \cite{pdg}:
\begin{eqnarray}
&&B^{exp}_p =3m_u-M_p=\frac32 M_\rho-M_p \ ,
\nonumber \\ 
&&B^{exp}_{\Lambda^0}=2m_u+m_s-M_{\Lambda^0}=M_{K^*} + \frac12 M_\rho- 
M_{\Lambda^0}\ , 
\nonumber \\
&&B^{exp}_{\Lambda^+_c}=2m_u+m_c-M_{\Lambda^+_c}=M_{D^*} + 
\frac12 M_\rho-M_{\Lambda^+_c} \ ,
\nonumber \\
&&B^{exp}_{\Lambda^0_b}=2m_u+m_b-M_{\Lambda^0_b}=M_{B^*} + 
\frac12 M_\rho-M_{\Lambda^0_b} \ .
\label{bind}
\end{eqnarray}
The results are presented in Table II. 

In figure 3, the binding energies of the low-lying pseudo-scalar mesons ($B_M$),
defined as the difference between the vector and pseudo-scalar masses (see
 Table I) are plotted against the mass of the corresponding pseudo-scalar
meson. Also, the values of the binding energies of the spin 1/2 baryons 
($N$, $\Lambda^0$, $\Lambda^+_c$ and $\Lambda^0_b$) from Eq.(\ref{bind}) are 
shown as a function of the corresponding baryon mass.
We observe  a smooth  trend in the plot of figure 3 for
$B_M$ as well as for $B_B$
as a function of the hadron ground state mass. The increase
of the heavy quark mass produces the same qualitative
behaviour irrespective of the nature of the hadron, being a meson or a 
baryon.

The data for mesons shown in figure 3, was described by
the effective QCD-inspired
model once the hypothesis of the flavor independence of the
interaction was adopted\cite{tobpauli}. This is consistent
with the fundamental QCD-theory in which
 the gluon does not recognize flavor but color\cite{iz}. 
Figure 3 suggests that
for baryons, it is reasonable to assume, as a first guess, that 
the constituent quark interaction would be flavor independent.

\subsection{Model calculations}

The physical input  of the light-front model
defined by the coupled integral equations (\ref{Mass3}) and (\ref{Mass32}) 
are the constituent quark masses and the diquark bound state mass.
Up to this point the masses of the constituent quarks have been defined by
Eq. (\ref{mconst}). The value of the diquark mass  has to be found.
The diquark mass was fitted to the value of the
proton mass, with a given constituent quark mass by the solution of 
Eq.(\ref{Mass3}) with $q=Q=u$. We use $m_u$= 0.386 GeV which together 
with the nucleon mass of 0.938 GeV implies in $M_d=$ 0.695 GeV. The slightly 
different $m_u$ in respect to the one found in Eq.s (\ref{mconst})
 is just to adequate to the value obtained in  Ref.\cite{Pacheco95}, 
where the above quark mass  was used with a reasonable 
description of the proton charge radius and 
electric form factor below 2(GeV/c)$^2$. 

We solve the coupled equations (\ref{Mass3}) and (\ref{Mass32}) for fixed
$m_u=0.386$ GeV and $M_d=$0.695 GeV and different values of $m_Q$. Thus,
the binding energy for the spin 1/2 baryons $\Lambda^0$, $\Lambda^+_c$
and $\Lambda^{0}_{b}$ are obtained by changing the value of $m_Q$ ($Q=s,c,b$).
Each value of $m_Q$ produces a ground state mass and binding energy.
In figure 4, instead of showing the binding energy
as a function of $m_Q$,  we plot it as a function of the ground state
mass in a continuous curve  and compare 
with the attributed experimental values. For the baryon mass above 
2.3 GeV, the  bound $Qqq$ system goes to the diquark threshold. This
gives the saturation value of $0.077 GeV$ seen in figure 4. 
The model calculation with a flavor independent effective interaction 
is able to reproduce the trend of the attributed experimental binding
energies as a function of the mass of the baryon ground state.
In view of the simplicity of the model,
the agreement between the binding energies obtained theoretically 
with the attributed experimental values is quite reasonable. 

Our present results, although, in a simplified model generalizes the flavor
independent effective interaction of the QCD-inspired $\uparrow\downarrow$-model 
to the context of the dynamics of constituent quarks forming the baryon. 
The very existence of the smooth pattern shown in figures 3 and 4, for the 
correlation between binding and masses, tells us that the dominant physics 
is the mass variation of the constituent quark and the spin effects should 
 average out. Taking into account that spin degree 
of freedom is averaged out in the model and at the same
time the reasonable description of the data found in figure 4,
give us the confidence that the main physics
related to quark mass variation is reasonable described by the 
flavor independent contact interaction.

\section{Conclusions}

We have studied the binding of the constituent quarks forming the
the low-lying spin 1/2 baryonic states of the nucleon, $\Lambda^0$, 
$\Lambda_c^+$ and $\Lambda_b^0$. We use a relativistic 
three-quark model of the baryon defined on the light-front, where
the inputs were the constituent quark masses and the diquark mass
in the light sector. The effective interaction was chosen of a contact
form and spin was averaged out. The contact interaction  
includes the minimal number of physical scales to describe these baryons.
Recently in a QCD-inspired model of mesons, besides the contact interaction,
motivated by the hyperfine interaction between the quarks, a Coulomb-like
potential was considered as well. Going beyond the hyperfine interaction
itself,
the zero range interaction mimics those aspects of QCD that  
binds the constituent quarks in the meson, which in this paper
was used to build the baryon. We observed a surprising reproduction
of the trend and magnitude of the binding energies as a function of the
distint quark mass. Our conclusion, while unexpected, still cares of
more detailed analysis which includes the quark spin and the Coulomb-type 
interaction.  The results shown here, give a  support for the
extension of the QCD-inspired model of Refs. \cite{pauli,tobpauli} to
baryons. 

As the feature we addressed here are valid irrespective of the
composite hadron nature, we think that our conclusion
of the flavor dependence of baryonic masses may still hold 
in a more realistic model. However, we have to stress that the present 
model is based on the notion of 
constituent quarks, and it only uses the constituent quark mass which in fact,
because spin effects are averaged, 
does not really enter into spin dependent interactions in our
calculations. When spin dependent interactions are present, 
constituent quark models 
can lead to conflict with data, as has been 
shown  in Ref.\cite{wei} for the proton spin measured in deep inelastic 
scattering. Therefore, the extension of our conclusion for the case that spins
are no longer averaged has to be cautious to avoid conflict
with spin data.

Acknowlegments: EFS and JPBCM thank the Brazilian funding agencies FAPESP 
(Funda\c{c}\~{a}o de Amparo a Pesquisa do
Estado de S\~{a}o Paulo) and TF thanks FAPESP and CNPq (Conselho Nacional de
Pesquisa e Desenvolvimento of Brazil).

\newpage
\appendix
\section{ Two-quark scattering amplitude}

The two-body scattering amplitudes $\tau _{qq}\left(M_{qq}^{2}\right)$ 
and $\tau _{Qq }\left( M_{Qq}^{2}\right)$
are the solutions of the 
Bethe-Salpeter equations in the ladder approximation,
represented diagrammatically in figure 5,  for
a contact interaction between the quarks \cite{tob92,Pacheco2001}.
In this case the solution is just given by 
the infinite sum of the product of "bubble"-diagrams (figure 6) multiplied
by powers of the bare interaction strength. The result is given
by the geometrical series:
\begin{equation}
\tau_{\alpha q} \left( M^2_{\alpha q}\right) =
\frac{1}{i\lambda^{-1}-{\cal B}_{\alpha q}\left( M^2_{\alpha q}\right)} \ ,
\label{Mass8}
\end{equation} 
where $\alpha = \ q$ or $Q$ and $\lambda$ is the bare interaction strength. 
The function ${\cal B}_{\alpha q}\left( M^2_{\alpha q}\right)$ is 
the "bubble"-diagram represented in figure 6,
\begin{equation}
{\cal B}_{\alpha q}\left( M^2_{\alpha q}\right) =\int \frac{d^4k}{\left( 2\pi 
\right)^4}\frac{i}{(k^{2}-
m_{q}^{2}+i\varepsilon) }\frac{i}{(( P_{\alpha q}-k)
^{2}-m_{\alpha}^{2}+i \varepsilon)}  \ ,  
\label{Mass4}
\end{equation} 
where the total four-momentum of the quark pair is $P_{\alpha q}$
with $P_{\alpha q}^2=M_{\alpha q}^2$. 

The four-dimensional integration in Eq.(\ref{Mass4}) is performed
in light-front variables. First, the
virtual propagation of the intermediate quarks  is projected at 
equal light-front times\cite{tob92,sales00}, by analytical 
integration over  $k^-$ in the momentum loop.
The non-zero contribution to Eq. (\ref{Mass4}) comes from
$0<k^{+}<P_{\alpha q}^{+}$ for $P^+_{\alpha q} > 0$, 
\begin{equation}
{\cal B}_{\alpha q}\left( M_{\alpha q}^2\right) 
=\frac{i}{2\left( 2\pi \right) ^{3}}
\int\frac{dk^{+}d^{2}k_{\perp }}{k^{+}\left(
P_{\alpha q}^{+}-k^{+}\right) }\frac{
\theta \left( P_{\alpha q}^{+}-k^+\right)\theta \left(k^+\right)}
{P_{\alpha q}^{+}-\frac{k_{\perp }^{2}-m_{q}^{2}}{k^{+}}-
\frac{\left( P_{\alpha q}^{+}-k\right) _{\perp }^{2}+m_{\alpha}^{2}}{\left(
P_{\alpha q}^{+}-k^{+}\right) }}.  \label{Mass5}
\end{equation}

Now, we introduce the invariant quantity $x=\frac{k^{+}}{P_{\alpha q}^{+}}$ 
and the relative momentum 
$$\vec{K}_{\perp}=(1-x)\vec{k}_{\perp}- 
x(\vec{P}_{\alpha q}-\vec{k}) _{\perp}\ ,$$
in Eq. (\ref{Mass5}), which gives:
\begin{equation}
{\cal B}_{\alpha q}\left( M^2_{\alpha q}\right)=
\frac{i}{2\left( 2\pi \right) ^{3}}\int
\frac{dxd^{2}K_{\perp }}
{x\left( 1-x\right) }\frac{\theta (1-x)\theta(x) }{M_{\alpha q}^{2}-\frac{
K_{\perp }^{2}+
\left( m^2_{\alpha}-m^2_{q}\right) x+m_{q}^{2}}{x\left(
1-x\right) }}.  
\label{Mass7}
\end{equation}

The function 
${\cal B}_{\alpha q}\left( M^2_{\alpha q}\right)$ has a log-type divergence
in the transverse momentum integration.
The renormalization $\tau_{\alpha q}(M^2_{\alpha q})$ is done 
by taking into account the 
physical information of the interacting light-quark pair system,
that we suppose has a bound state. Using this physical condition 
to define the two-quark scattering amplitude we have studied
the nucleon in the three-quark light-front model\cite{Pacheco95}.
This model fitted,  simultaneously, the  proton mass, the
charge radius and the electric form factor below 2 (GeV/c)$^2$\cite{Pacheco95}. 
Here, we just use the same renormalization condition. 

The pole of the light-quark 
scattering amplitude, $\tau_{qq}(M^2_{qq})$ is 
found when $M_{q q}$ is equal to the mass of the bound $qq$ pair, $M_{d}$.
Thus, the bound state pole of the scattering amplitude demands that
\begin{equation}
i\lambda ^{-1}={\cal B}_{qq}\left( M^2_{d}\right) \ ,
\label{Mass9}
\end{equation}
which is enough to render finite the scattering amplitudes $\tau_{qq}$
and $\tau_{Qq}$. Thus, the bare strength of the
effective contact interaction between the constituent quarks $q$ and
$Q$ does not depend on flavor. In this manner, we extend the
flavor independence of the gluon interaction of the fundamental QCD
Lagrangian to the effective interaction.

The final  equation for the two-quark scattering amplitude is:
\begin{equation}
\tau_{\alpha q} \left( M^2_{\alpha q}\right) =
\frac{1}{{\cal B}_{qq}\left( M^2_{d}\right)-
{\cal B}_{\alpha q}\left(M^2_{\alpha q}\right)} \ .
\label{Mass11}
\end{equation} 
The log-type divergence of $\tau_{\alpha q}$ is removed by the subtraction in
Eq.(\ref{Mass11}).

In particular,  the analytical form of Eq.(\ref{Mass11}) is \cite{tob92}, 
\begin{eqnarray}
\tau _{qq }\left( M_{qq}^{2}\right) &=&-i\left( 2\pi \right)
^{2}\left\{ \sqrt{\frac{m_q^{2}}{M_{d}^{2}}-\frac{1}{4}}
\arctan \left( \frac{%
1}{2\sqrt{\frac{m_q^{2}}{M_{d}^{2}}-\frac{1}{4}}}\right) 
\right. -  \nonumber
\\
&&\left. \sqrt{\frac{m^{2}_q}{M_{qq}^{2}}-\frac{1}{4}}
\arctan \left( \frac{1}{2%
\sqrt{\frac{m^{2}_q}{M_{qq}^{2}}-\frac{1}{4}}}\right) \right\} ^{-1} \ ,
\label{Mass12}
\end{eqnarray}
for $0 < M_{qq}<2m_{q}$, which is enough for the integration
in  Eq.(\ref{Mass3}).



\newpage

\begin{table}

\caption{ Vector (pseudo-scalar) meson and  current quark masses
 from Ref. [16]. The estimated values of the current quark masses
(last column) are explained in the text. Values quoted in GeV. }

\begin{tabular}{cccccc}
Meson &    & $ M_M$    & quark  &$m^{curr}_q$ (Ref. [16]) &  $ m^{curr}_q $  \\ \hline 
$\rho \ (\pi) $ &$u\overline d$ & 0.768 (0.138) &$u,d$& (1.5 - 9)$\times 10^{-3}$& -
\\
$K^* \  (K^+) $ &$u\overline s$ & 0.892 (0.494) & $s$ &0.06 - 0.17 & 0.124 
\\ 
$D^* \ (D^0)$ &$\overline u c$ & 2.007 (1.865)   & $c$ & 1.15 - 1.30 & 1.115  
\\ 
$B^* \ (B^+)$ &$u\overline b$ & 5.325 (5.279)   & $b$ & 4 - 4.4 &  4.557  
\end{tabular}
\end{table}  

\begin{table}

\caption{ Low-lying spin 1/2 baryon experimental masses from Ref. [16] 
and binding energies from Eq. (\ref{bind}). Values quoted in GeV. }

\begin{tabular}{ccccc}
Baryon       &               & $I(J^P)$       &$M_{B}$ &$B_B$  
\\ \hline
$p$            &     $uud$ & $ \frac12(\frac12^+)$ & 0.938 & 0.214 
\\ 
$\Lambda^{0}$  &     $uds$ & $0(\frac{1}{2}^+)$    & 1.115 & 0.161 
\\ 
$\Lambda _{c}^{+}$  &$udc$ & $0(\frac{1}{2}^+)$    & 2.285 & 0.106  
\\
$\Lambda _{b}^{0}$  &$udb$ & $0(\frac{1}{2}^+)$    & 5.624 & 0.085   
\end{tabular}
\end{table}

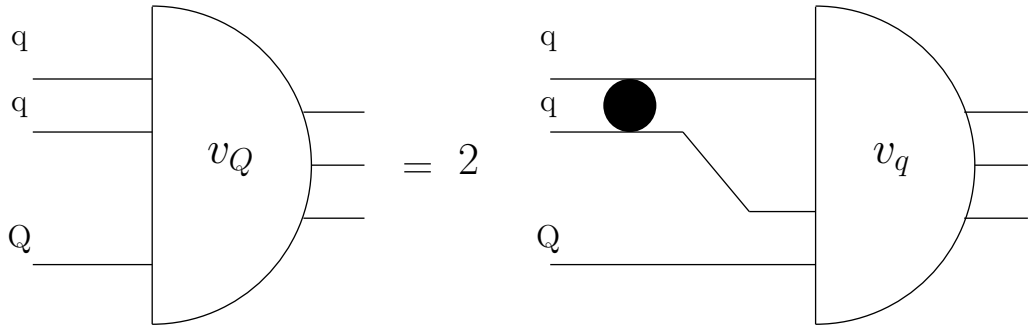
\begin{figure}[h]

\begin{center}

\begin{picture}(330,200)(0,0)

\CArc(50,117.5)(60,270,90)

\begin{Large}

\Text(80,120)[]{$v_Q$}
\end{Large}

\Line(50,57.5)(50,177.25)

\Text(0,90)[]{Q}

\Text(0,140)[]{q}

\Text(0,165)[]{q}

\Line(5,150)(50,150)

\Line(5,130)(50,130)

\Line(5,80)(50,80)

\Line(110,117.5)(130,117.5)

\Line(107,137.5)(130,137.5)

\Line(107,97.5)(130,97.5)

\Line(145,118)(155,118)

\Line(145,115)(155,115)

\begin{Large}

\Text(170,120)[]{2}

\end{Large}

\Text(200,90)[]{Q}

\Text(200,140)[]{q}

\Text(200,165)[]{q}

\CArc(300,117.5)(60,270,90)

\begin{Large}

\Text(330,120)[]{$v_q$}
\end{Large}
\Line(300,57.5)(300,177.25)

\Line(200,150)(300,150)

\Line(200,130)(250,130)

\Line(250,130)(275,100)

\Line(275,100)(300,100)

\Line(200,80)(300,80)

\Vertex(230,140){10}

\Line(360,117.5)(380,117.5)

\Line(356,137.5)(380,137.5)

\Line(356,97.5)(380,97.5)

\end{picture}

\end{center}

\caption{Diagrammatic representation of Eq. (1). The black bubble represents
the two-quark scattering amplitude.}

\label{fig.1}

\end{figure}

\vskip 2cm

\begin{figure}[h]

\begin{picture}(330,200)(0,0)

\CArc(50,117.5)(60,270,90)

\begin{Large}

\Text(80,120)[]{$v_q$}
\end{Large}
\Line(50,57.5)(50,177.25)

\Text(0,90)[]{q}

\Text(0,140)[]{q}

\Text(0,165)[]{Q}

\Line(5,150)(50,150)

\Line(5,130)(50,130)

\Line(5,80)(50,80)

\Line(110,117.5)(130,117.5)

\Line(107,137.5)(130,137.5) 

\Line(107,97.5)(130,97.5)

\Line(145,118)(155,118)

\Line(145,115)(155,115)

\Text(170,90)[]{q}

\Text(170,140)[]{q}

\Text(170,165)[]{Q}

\CArc(240,117.5)(60,270,90)

\begin{Large}

\Text(270,120)[]{$v_q$}
\end{Large}
\Vertex(200,140){10}

\Line(240,57.5)(240,177.25)

\Line(170,150)(240,150)

\Line(170,130)(215,130)

\Line(215,130)(230,97.5)

\Line(230,97.5)(240,97.5)

\Line(170,80)(240,80)

\Text(220,165)[]{q}

\Text(220,140)[]{Q}

\Line(297,137.5)(320,137.5)

\Line(300,117.5)(320,117.5)

\Line(297,97.5)(320,97.5)

\begin{Large}

\Text(335,110)[]{+} 

\end{Large}

\Text(340,90)[]{q}

\Text(340,140)[]{q}

\Text(340,165)[]{Q}

\CArc(410,117.5)(60,270,90)

\begin{Large}

\Text(440,120)[]{$v_Q$}
\end{Large}
\Line(410,57.5)(410,177.25)

\Line(340,150)(410,150)

\Line(340,130)(380,130)

\Vertex(360,140){10}

\Line(380,130)(400,97.5)

\Line(400,97.5)(410,97.5)

\Line(340,80)(410,80)

\Line(467,137.5)(490,137.5)

\Line(470,117.5)(490,117.5)

\Line(467,97.5)(490,97.5)

\Text(380,140)[]{q}

\Text(380,165)[]{Q}

\end{picture}

\caption{Diagrammatic representation of Eq. (2). The black bubbles represent
the two-quark scattering amplitudes. }
\label{fig.2}

\end{figure}
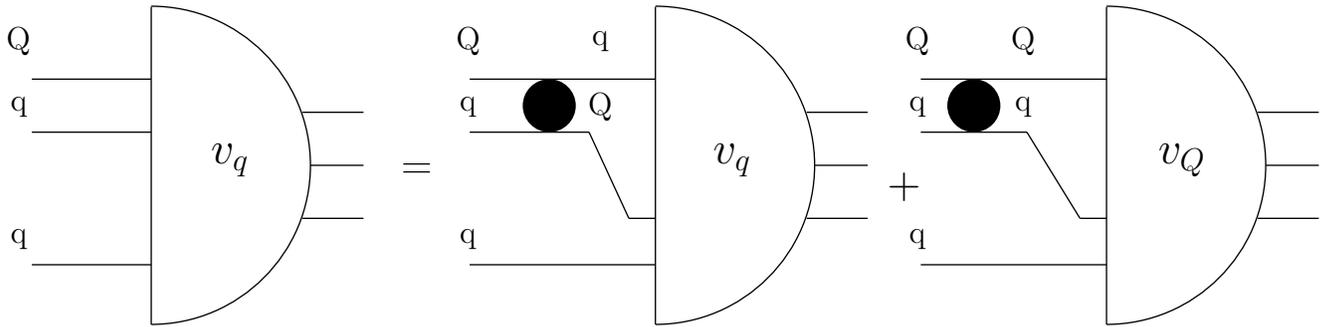

\newpage 
.
\begin{figure}[h]
\includegraphics{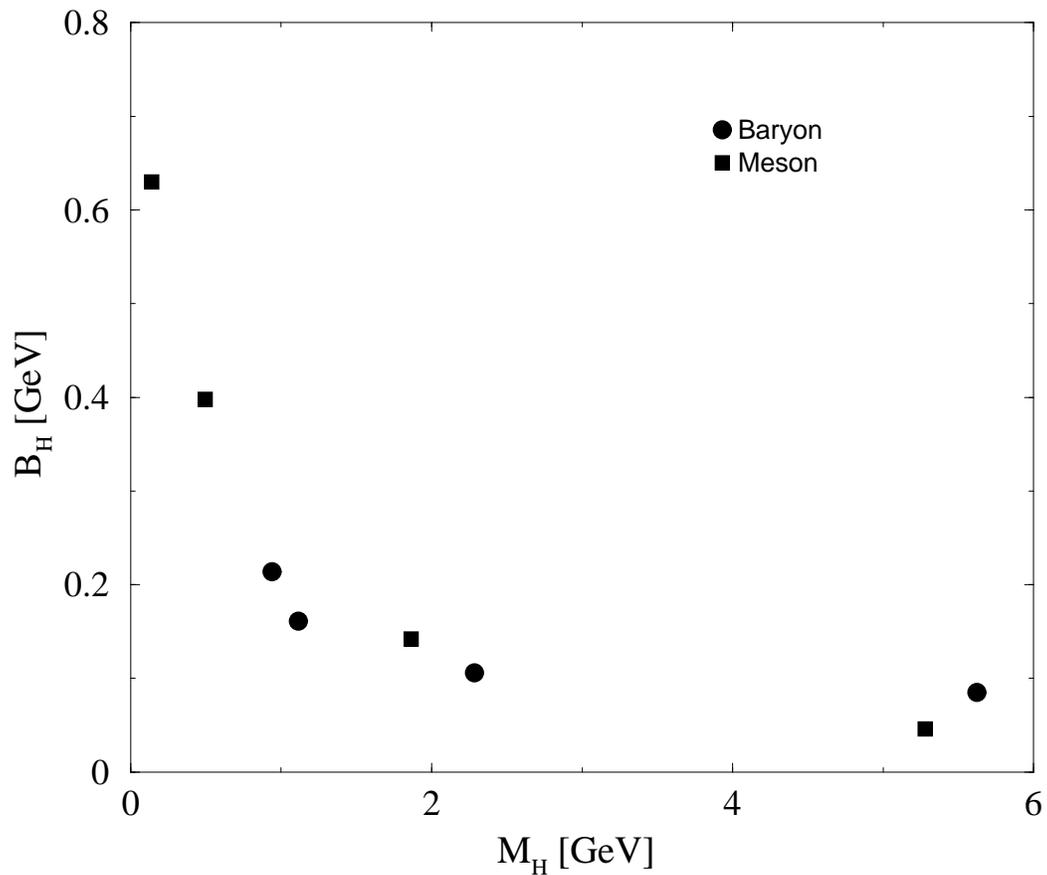}
\vspace{15.0cm}
\caption{Binding energy $(B_H)$ as a function of the mass of the respective 
low-lying hadron ($M_H$). Experimental data of pseudo-scalar mesons from Table I 
(full squares). Experimental data of the spin 1/2 baryons from Table II
(full circles).}
\label{fig.3}
\vspace{1.0cm} 
\end{figure}  

\newpage  
.
\begin{figure}[h]
\vspace{5.0cm}
\includegraphics{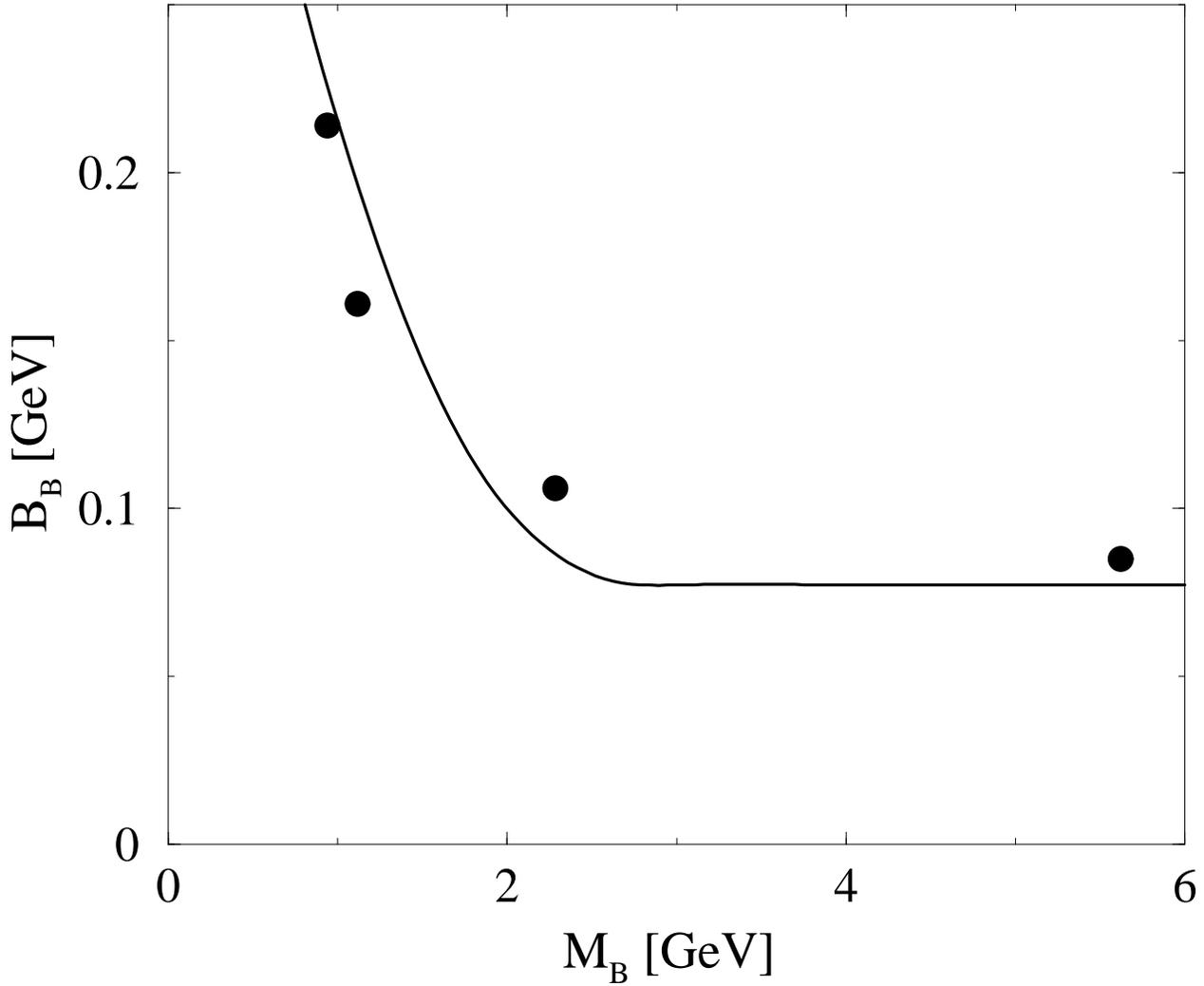} 
\vspace{13.0cm}
\caption{Binding energy of the low-lying spin 1/2 baryon states
as a function of  the respective ground state mass of the baryon. The results
of the light-front Faddeev model calculation are shown by the solid line.
Full circles are the  data  from Table II for the nucleon, 
$\Lambda^{0}$, $\Lambda^{+}_{c}$ and $\Lambda^{0}_b$. }
\label{fig.4}
\vspace{1.0cm} 
\end{figure}

\newpage

\begin{figure}[h]

\begin{center}

\begin{picture}(330,50)(0,0)

\Line(10,40)(80,40)

\Line(10,0.0)(80,0.0)

\GOval(40,20)(10,20)(90){0}

\Line(110,25)(115,25)

\Line(110,20)(115,20)

\Line(130,40)(180,0.0)

\Line(180,40)(130,0.0)

\Vertex(155,20){5}

\begin{Large}

\Text(210,20)[]{+}

\Line(250,40)(280,20)

\Line(250,0.0)(280,20)

\Vertex(280,20){5}

\CArc(300,20)(20,90,270)

\Line(300,40)(350,40)

\Line(300,0.0)(350,0.0)

\GOval(320,20)(10,20)(90){0}

\end{Large}

\end{picture}

\end{center}

\caption{Bethe-Salpeter equation in ladder approximation for the 
two quark-scattering amplitude from Eq. (\ref{Mass8}). 
The contact interaction is represented by the dot.}

\label{fig.5}

\end{figure}
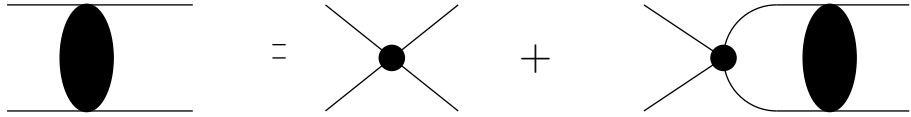

\vspace{2cm}

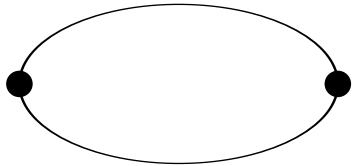
\begin{figure}[h]

\begin{center}

\begin{picture}(330,80)(0,0)

\GOval(170,80)(30,60)(0){1}

\Vertex(110,80){5}

\Vertex(230,80){5}

\end{picture}

\end{center}

\caption{"Bubble"-diagram representing Eq.(\ref{Mass4}).}

\label{fig.6}

\end{figure}


\begin{thebibliography}{99}

\bibitem{iz} C. Itzykson and J.-B. Zuber, Quantum Field Theory, McGraw-Hill (1987).

\bibitem{brodsky} S.J. Brodsky, H.-C. Pauli, and S. S. Pinsky,
Phys. Rep. {\bf 301}, 299 (1998).

\bibitem{karmanov} J. Carbonell, B. Desplanques, V. Karmanov, J.-F. Mathiot,
Phys. Rep. {\bf 300}, 215 (1998).

\bibitem{perry}  R.J. Perry, A. Harindranath, K.G. Wilson, Phys. Rev. Lett. 
{\bf 65}, 2959 (1990).

\bibitem{pauli} H.-C. Pauli, in: New Directions in Quantum Chromodynamics, C.R.Ji
and D.P. Min, Eds., AIP(1999) 80-139;
Nucl. Phys. {\bf B} (Proc. Suppl.) {\bf 90}, 154 (2000); ibid., 259 (2000).

\bibitem{tobpauli} T. Frederico, H.-C. Pauli., Phys. Rev. {\bf D64:}
054007 (2001).

\bibitem{tob92}T. Frederico, Phys.Lett. {\bf B282}, 409 (1992); S.K.
Adhikari, L. Tomio and T. Frederico, Ann. Phys. {\bf 235}, 77 (1994).

\bibitem{Pacheco95}W.R.B. Ara\'{u}jo, J.P.B.C. de Melo, T. Frederico, Phys.
Rev. {\bf C52}, 2733 (1995).

\bibitem{pauli98} H.-C. Pauli, Eur. Phys. J. {\bf C7}, 289 (1998).

\bibitem{beyer01} S. Matiello, M. Beyer, T. Frederico, H.-J. Weber,
Phys. Lett. {\bf B251}, 33 (2001).

\bibitem{Weinberg66}S. Weinberg, Phys. Rev. {\bf 150}, 1313 (1966).

\bibitem{sales00} J.H.O. Sales, T. Frederico, B.V. Carlson,
P.U. Sauer, Phys. Rev. {\bf C61:}044003 (2000); ibid., {\bf C63:}064003
 (2001).

\bibitem{miller} J. R. Cooke, G. A. Miller, D. Phillips, Phys. Rev.
{\bf C61:}064005 (2000).

\bibitem{brinet} M. Mangin-Brinet and J. Carbonell, Phys. Lett. {\bf B474},
237 (2000).

\bibitem{Pacheco2001}J.P.B.C. de Melo, A. E. A. Amorim, 
L. Tomio and T. Frederico, J. Phys. G {\bf 27}, 1031 (2001).

\bibitem{pdg}Particle Data Group, Eur. Phys. J. {\bf C15}, 1 (2000).

\bibitem{wei} K. Suzuki and W. Weise, Nucl.Phys. {\bf A634}, 141 (1998).

\end{thebibliography}
\end{document}